\documentstyle[12pt,epsf]{article}

\def\beq{\begin{equation}}
\def\eeq{\end{equation}}
\def\beqn{\begin{eqnarray}}
\def\eeqn{\end{eqnarray}}
\def\q{\quad}

\begin{document}

\begin{titlepage}
\begin{flushright}
       {\normalsize
          hep-th/9703064 \\
          OU-HET 260 \\
          March 1997 \\}
\end{flushright}
\vfill
\begin{center}
  {\large T-Duality and Time } \\
  {\large development of a (2+1)-Dimensional } \\
  {\large  String Universe } \\
\vfill
           {\bf Kenji Hotta, Keiji Kikkawa and Michihiko Sakamoto } \\
\vfill
        Department of Physics,\\
        Graduate School of Science, Osaka University,\\
        Toyonaka, Osaka, 560 Japan\\

\end{center}
\vfill

\begin{abstract}

The time development of a model of (2+1)-dimensional torus universe is studied based on background field equations which follow from a string theory. The metrics in various cases are characterized by a real parameter which specifies a ratio of the lengths of two independent cycles. When the parameter is a rational number, the space is asymptotically stretched along a cycle while the other cycle kept finite. When the parameter is an irrational number, the lengths of two cycles, as well as the space volume (area), grow in proportion to the proper time $t$ for an observer sitting at rest in this universe in the asymptotic region. 
\end{abstract}
\vfill
\end{titlepage}

\section{Introduction}

If string theories in critical dimensions are applied to cosmology, the stability of the small compactified space other than (3+1)-dimensions is required in the asymptotic time region. In a superstring model it was previously pointed out that the radii of those shrunken space are stable and should be of the order of the Planck size provided there is existence of small supersymmetry violation $\cite{kikkawa}$. There T-duality played a critical role.

In this article, we study the time developments of various solutions for the torus universe in (2+1)-dimensions and look for a possibility of partial shrinkage of the space in the asymptotic region by solving the equations of motion which follow from the low energy effective action of the string theory. The crucial points of our arguments again rest on the T-duality.

To study the quantum states of string, other than string oscillation modes, on a compactified space, Brandenberger and Vafa $\cite{brandenberger}$ introduced a new space coordinate $\tilde{x}$. If the space is compactified as a circle $S^{1}$, for example, one has to consider two sets of quantum numbers for non-oscillation modes. One is the center-of-mass momenta $p$ on $S^{1}$ and the other is the winding number $w$ of a string wound around the space $S^{1}$, both are chosen as integers under a certain scale convention. Associated with the momentum state $|\;p\,>$, the coordinate diagonalized state $|\;x\,>$ is defined by
\beq
         |\; x\,>=\sum_{p}\exp (ix\cdot p)|\; p\,>.
\eeq
Analogously one can introduce the new space corresponding to the winding number diagonalized state $|\;w\,>$ by
\beq
         |\:\tilde{x}\,>=\sum_{w}\exp (i\tilde{x}\cdot w)|\: w\,>.
\label{eq2}
\eeq
In this article the space spanned by (\ref{eq2}) is called $\tilde{x}$-space. Just as the center-of-mass coordinate $x^{\mu}$ is given by the zero frequency mode of the string field $X^{\mu}(\sigma,\;\tau)$, the new coordinate $\tilde{x}^{\mu}$ is given by the zero frequency mode of the T-transformed field $\tilde{X}^{\mu}(\sigma,\;\tau)$.

Taking into account these $x$ and $\tilde{x}$ spaces, the time development of background metric was studied by Tseytlin and Vafa $\cite{tseytlin}$ under the low energy effective action provided the metric being diagonal. Gasperini and Veneziano, and other people $\cite{veneziano}$, studied the same problem in connection with the inflational universe but there again the off-diagonal elements of metric tensor were disregarded. As far as we know the problem has not been studied with the effect of off-diagonal elements.

In this article we treat a model of bosonic string in (2+1)-dimensional space. The arguments are, however, applicable for the critical strings provided that the two-dimensional space is separated by a direct product from the rest of space which is assumed to be flat and infinitely extended.

We find that the asymptotic forms of the space are characterized by the number $\sqrt{{G_{11} \over G_{22}}}$ in $t\rightarrow\infty$ limit, where $G_{ij}$ represent spatial components of the metric tensor.


\section{(2+1)-dimensional cosmological solutions}

The model we concern is a bosonic closed string theory on general background fields and the action is given by
\beqn
  S &=&{1 \over 4{\pi}{\alpha}'}\int d^{2}{\sigma}[\sqrt{h}h^{\alpha\beta}
      g_{\mu \nu}(X)\partial_{\alpha}X^{\mu}\partial_{\beta}X^{\nu}
\nonumber\\
      &\qquad+&\varepsilon^{\alpha\beta}B_{\mu\nu}(X)\partial_{\alpha}X^{\mu}
      \partial_{\beta}X^{\nu}-\alpha'\sqrt{h}R^{(2)}\Phi(X)],
\label{eq3}
\eeqn
where $\alpha'$ is the slope parameter. The fields $g_{\mu\nu}$, $B_{\mu\nu}$ and $\Phi$ represent respectively the background metric, the antisymmetric tensor and the dilaton. The conformal invariance of the theory, i.e. the vanishing condition of $\beta$-functions provides us the equations for background fields
\beqn
      R_{\mu\nu}-{1 \over 4}H_{\mu}^{~\lambda\rho}H_{\nu\lambda\rho}
                +2D_{\mu}D_{\nu}\Phi&=&0 
\label{eq4}\\
      D_{\lambda}H^{\lambda}_{~\mu\nu}-2(D_{\lambda}\Phi)H^{\lambda}_{~\mu\nu}&=&0 
\label{eq5}\\
      4(D_{\mu}\Phi)^{2}-4D_{\mu}D^{\mu}\Phi-R
                        +{1 \over 12}H_{\mu\nu\rho}H^{\mu\nu\rho}&=&0,
\label{eq6}
\eeqn
where
\beqn
      H_{\mu\nu\rho}=\partial_{\mu}B_{\nu\rho}+\partial_{\rho}B_{\mu\nu}
                     +\partial_{\nu}B_{\rho\mu}.
\nonumber
\eeqn
These equations are valid up to $O(\alpha')$ and derived from the low energy effective action. The physics we concern is the (2+1)-dimensional compactified space-time. The boundary conditions over strings for the compactified components are chosen for dimensionless coordinates $X^{i}$ as
\beq
         X^{i} \approx X^{i}+2\pi m^{i}\q,\qquad i\;=\;1,\;2,
\label{eq7}
\eeq
where $m^{i}$ is called winding number and takes values 0, $\pm1,\:\pm2,\cdots\;$. The space-time metric is chosen as
\beq
      g_{\mu\nu}=\left( \begin{array}{cc}
             -1 & 0 \\
              0 & e^{\lambda(t)}\tilde{g}_{ij}(t)
             \end{array} \right) 
\eeq
\beq
      G_{ij}=e^{\lambda(t)}\tilde{g}_{ij}(t)
\label{eq9}
\eeq
\beqn
      \det \tilde{g}_{ij}=1  \qquad          i,j=1,2.
\nonumber
\eeqn
Here the parameter $t$ represents the proper time for an observer sitting at rest in this universe. The antisymmetric field is defined as
\beqn
      B_{\mu\nu}= \left(\begin{array}{ccc}
                           0 & 0 & 0 \\
                           \begin{array}{c}
                            0 \\
                            0
                           \end{array} & ~~\qquad ~B_{ij} \\
                          \end{array}\right),
\nonumber
\eeqn
with
\beq
      B_{ij}= \left(\begin{array}{cc}
                     0 & \alpha(t) \\
                     -\alpha(t) & 0 \\
                   \end{array}\right).
\label{eq10}
\eeq

In solving the equations $(\ref{eq4})\sim(\ref{eq6})$, we assume the dilaton is a constant $\Phi_{0}$. However, if the solution is transformed by T-duality the transformed one is also a solution to the equation of motion, in which the dilaton usually shows time dependence. To understand characteristic behavior of solutions we first study a special case where the metric tensor $\tilde{g}_{ij}$ has common diagonal elements
\beq
        \tilde{g_{ij}}=\left( \begin{array}{cc}
                         \cosh \theta(t) & \sinh \theta(t) \\
                         \sinh \theta(t) & \cosh \theta(t) \\
                        \end{array} \right).
\label{eq11}
\eeq
Substituting this into the background equations $(\ref{eq4})\sim(\ref{eq6})$ one can find the solution I for $g_{\mu\nu}$ and $B_{ij}$ as follows.
\beq
   {\rm I}\q:\qquad g_{\mu\nu}=\left( \begin{array}{ccc}
                   -1  &  0  &  0\\
                    0  &  {1+t^{2} \over 2}  &  {-1+t^{2} \over 2}\\
                    0  &  {-1+t^{2} \over 2}  &  {1+t^{2} \over 2}
                    \end{array} \right) \qquad
        B_{ij}=\left( \begin{array}{cc}
                   0 & b \\
                  -b & 0 
                  \end{array}\right) ,
\label{eq12}
\eeq
with
\beq
        \sqrt{-g}=\; t \qquad \qquad \qquad b= const,
\eeq
where we so adjusted integration constants to make the expressions simple.

The generic solution for $G_{ij}=e^{\lambda(t)}\tilde{g}_{ij}(t)$ should have one more independent function $\beta(t)$ and
\beq
        \tilde{g_{ij}}=\left( \begin{array}{cc}
                      e^{\beta(t)} \cosh \theta(t) & \sinh \theta(t) \\
                      \sinh \theta(t) & e^{-\beta(t)} \cosh \theta(t) \\
                        \end{array} \right).
\label{eq14}
\eeq
Substituting (\ref{eq14}) into the equations of motion $(\ref{eq4})\sim(\ref{eq6})$ we are able to find the solutions II as follows.
\beq
   {\rm II}~:\quad g_{\mu\nu}=\left( \begin{array}{ccc}
                   -1  &  0  &  0\\
                    0  & e^{\beta}\sqrt{t^{2}+c^{2}\left({t^{2} \over 2}+d\right)^{2}} &  c\left({t^{2} \over 2}+d\right)\\
                    0  & c\left({t^{2} \over 2}+d\right) & e^{-\beta}\sqrt{t^{2}+c^{2}\left({t^{2} \over 2}+d\right)^{2}}
                    \end{array} \right)
\label{eq15}
\eeq
\beq
        B_{ij}=\left( \begin{array}{cc}
                   0 & b \\
                  -b & 0 
                  \end{array}\right),
\label{eq16}
\eeq
where
\beq
        e^{\beta}= f\sqrt{{t^{2}+2d+{2 \over c^{2}}-{2 \over c^{2}}\sqrt{1+2c^{2}d} \over t^{2}+2d+{2 \over c^{2}}+{2 \over c^{2}}\sqrt{1+2c^{2}d}}}
\eeq
\begin{center}
$\sqrt{-g}= t$ \qquad $b, c, d, f = const$ .
\end{center}

In order to describe the properties of the space structure and the T-duality of our model we use the following parameters
\beqn
        \tau&\equiv&{G_{12} \over G_{22}}+{i\sqrt{G} \over G_{22}}\\
        \rho&\equiv&B_{12}+i\sqrt{G}.
\eeqn
The space structure is, therefore, characterized by a set of two complex numbers
\beq
        (\;\tau,\;\rho\;),
\eeq
where $\tau$ and $\rho$ are, respectively, the moduli parameter and the {\it K$\ddot{a}$hler} structure. The general T-duality for toroidal compactifications has been well studied in literatures $\cite{giveon}$, and for the case of (2+1)-dimensions the duality is generated by four independent discrete transformations which are the elements of O(2,2,{\bf Z}) $\tau\rightarrow\tau+1,~\tau\rightarrow-{1 \over \tau},~\tau\leftrightarrow\rho$ and a world sheet parity $\rho\rightarrow-\bar{\rho}$. One of the dual transformation interchanges the parameters $\tau$ and $\rho$, hence $(\tau,\;\rho)$ and $(\rho,\;\tau)$ represent the spaces with identical spectrum. The time developments of our space structure will be demonstrated in $(\tau,\;\rho)$ in the following sections.

In general the T-duality transformations are defined in terms of the background matrix $\cite{giveon}$
\beq
        E\;=\;G\;+\;B,
\eeq
where G and B are the 2$\times$2 matrices introduced in (\ref{eq9}) and (\ref{eq10}), respectively. The duality transformation which plays a major role in our work is the one associated with the transformation
\beq
        E\q\longrightarrow\q E^{-1},
\label{eq22}
\eeq
which is a generalization of the radius $R\rightarrow R^{-1}$ for $S^{1}$-model, and interchanges the $x$-space and the $\tilde{x}$-space. The duality transformation (\ref{eq22}) is referred to as $T_{D}$.


\section{The Solution I}

\begin{center}
\leavevmode
\epsfysize=6cm \epsfbox{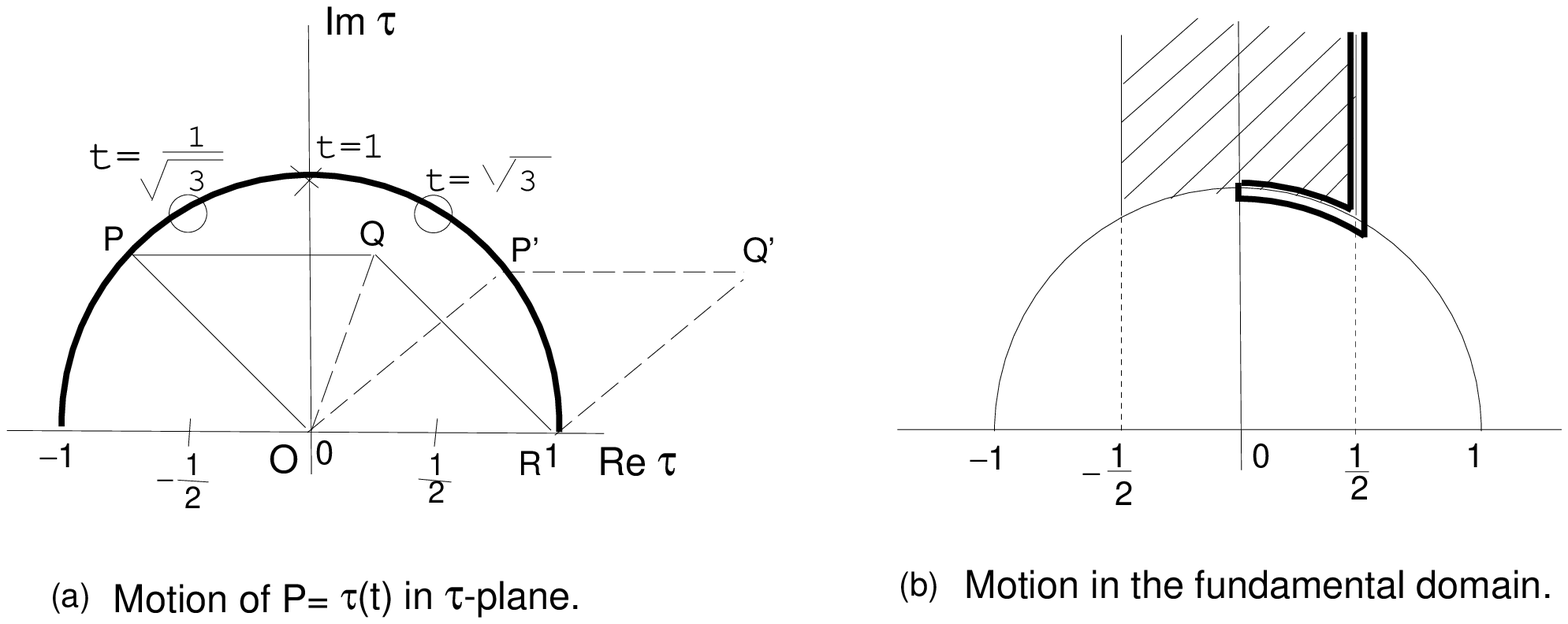}
\end{center}
\begin{center}
{\small fig.1 ~Motion of the moduli $\tau$.}
\end{center}

For the solution I with $b=0$, for instance, the motion in the space of $\tau$  for the time $t$ is represented with the point P on the parallelogram OPQR in fig.1-(a), and the solution of $\rho$ is represented with the point P in fig.2-(a). When the trajectory of $\tau$ and $\rho$ for the time $t$ are out of the fundamental region, we can transform the parameters into the fundamental region of $SL(2,Z)$ by moduli transformations. Then both $\tau$ and $\rho$ trace an identical point twice, except for the point $\tau=\rho=i$ in the fundamental region depicted in fig.1-(b) and fig.2-(b).

\begin{center}
\leavevmode
\epsfysize=6cm \epsfbox{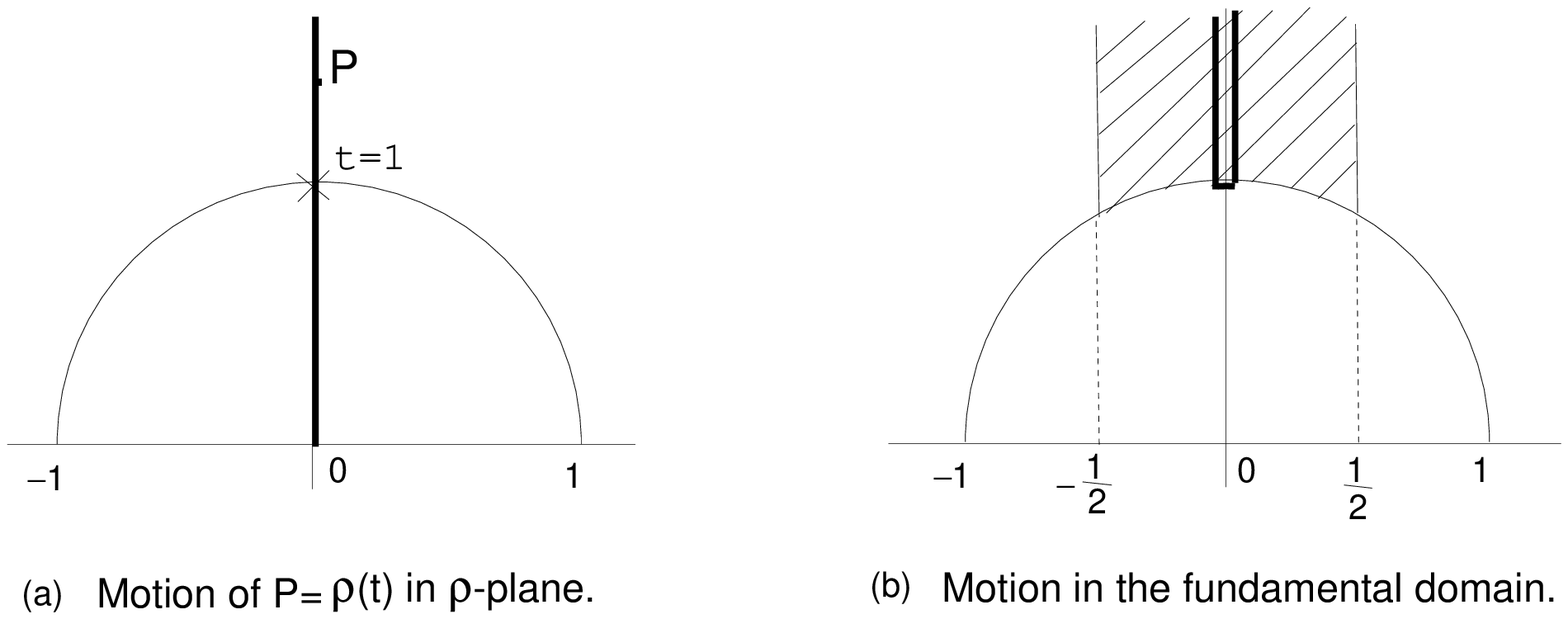}
\end{center}
\begin{center}
{\small fig.2 ~Motion of the {\it K$\ddot{a}$hler} structure $\rho$.}
\end{center}

We have another way of representing the structure of the torus, namely, representing the lengths of the two independent cycles of the torus. The line element on the torus $ds$ is given by
\beq
          (ds)^{2}=G_{ij}dX^{i}dX^{j},
\label{eq23}
\eeq
where G is defined in (\ref{eq9}). The variation range of the 1-cycle of $X^{i}$ has been normalized to be 2$\pi$ as in (\ref{eq7}).

For the solution I, using (\ref{eq12}) and (\ref{eq23}), the cycle length along $X^{2}$-direction $s_{2}$ is given by
\beq
          s_{2}^{~2}=2\pi^{2}(1+t^{2}).
\label{eq24}
\eeq
The other circumference along $X^{1}$-direction is chosen along the shortest cycle. In fig.1-(a) the shortest cycle $s_{1}$ is first defined along $\overline{OQ}$, but as time goes on, P moves along the semi-circle from left to right and at a certain point, the distance $\overline{OP}$ becomes equal to $\overline{OQ}$ and shorter after this point. Then the $s_{1}$ is redefined as the distance along $\overline{OP}$. Furthermore when P moves on, $\overline{PR}$ becomes even shorter and $s_{1}$ is measured along $\overline{PR}$. For the solution I, then, $s_{1}$ is represented as follows :
\beq
          s_{1}^{~2}=\left\{ \begin{array}{lc}
                               8\pi^{2}t^{2}, &  0<t\leq{1 \over \sqrt{3}}\\
                               2\pi^{2}(1+t^{2}), &  {1 \over \sqrt{3}}<t\leq \sqrt{3}\\
                               8\pi^{2}, &  \sqrt{3}<t.
                             \end{array} \right.
\label{eq25}
\eeq
The result is shown in fig.3.

\begin{center}
\leavevmode
\epsfysize=9cm \epsfbox{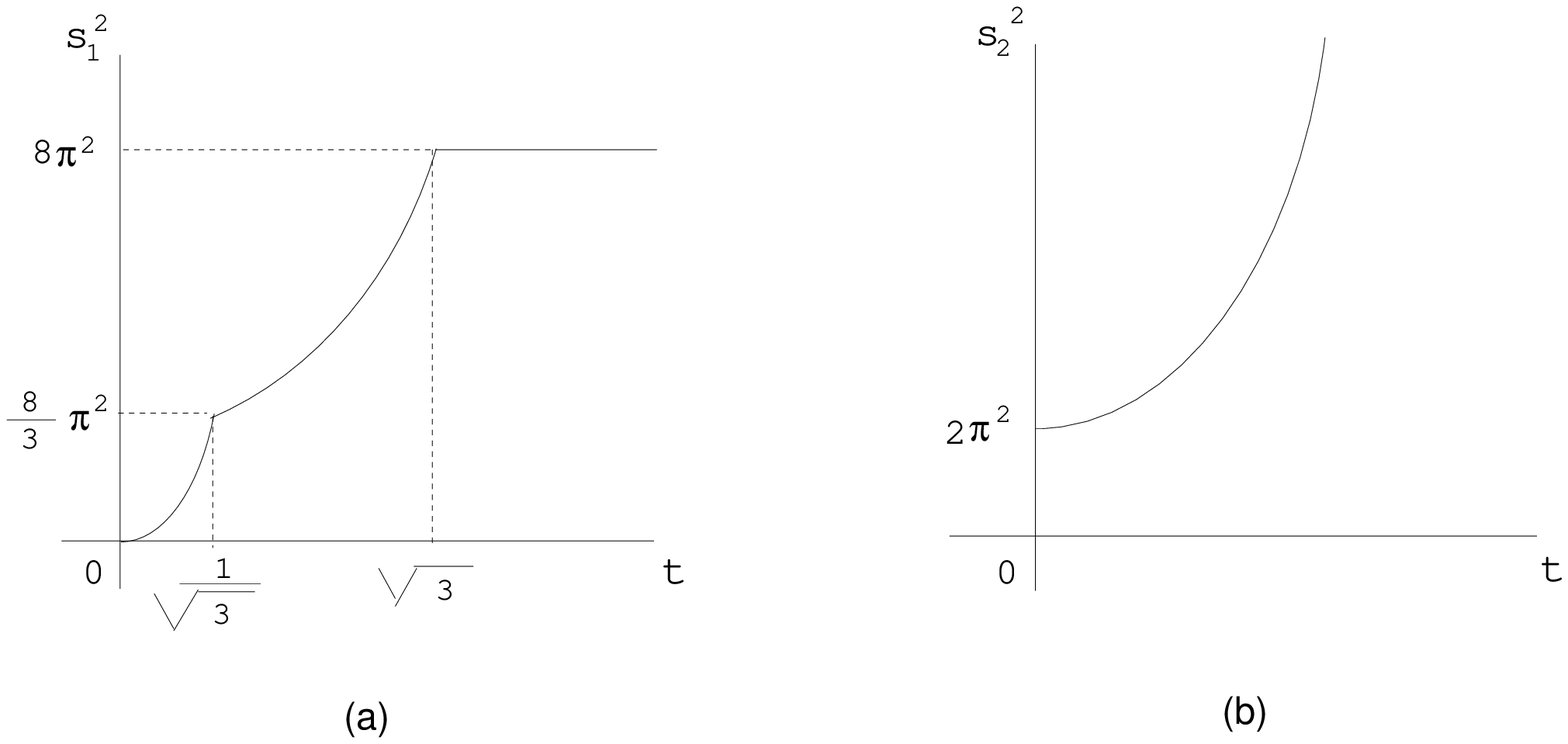}
\end{center}
\begin{center}
{\small fig.3 ~The cycle lengths in $x$-space for the solution I.}
\end{center}

The solution I shows special gauge symmetries such as $SU(2)\times SU(2)$ or $SU(3)$ at some particular time $\cite{giveon},~\cite{sakai}$. At the moment of $t=1$ both $\tau$ and $\rho$ take a pure imaginary value $i$ and there the system has $[SU(2)\times SU(2)]^{2}$ gauge symmetry, and at $t=(\sqrt{3})^{\pm 1}$, an $SU(3)$ gauge symmetry.

When the torsion $b\ne 0$, no essential difference appears but the behavior of the {\it K$\ddot{a}$hler} structure becomes different and the gauge symmetry at $t=1$ is reduced to $SU(2)\times SU(2)$.

If this compactified space is identified with a part of our space, the internal gauge symmetry changes as time goes on. These time-dependent symmetry, however, may not be observed in the asymptotic region. Because only the dimensionful scale factor in the model is the slope parameter $\alpha'$ which is of the order of the Planck scale, the time dependence appears only at an early stage $t\approx 1$ and none in the asymptotic region $t\rightarrow \infty$, where we are supposed to live.

Here some important comments are in order. First, the length of $s_{1}$-cycle goes to zero at $t=0$. But if one looks at the size of $T_{D}$-dual space, which means the interchanging of $\tau\rightarrow -{1 \over \tau}$ and $\rho\rightarrow -{1 \over \rho}$ in this case $\cite{giveon}$, the spatial components of the metric tensor turns out to be
\beq
           G_{D}=\left( \begin{array}{cc}
                           {1+t^{2} \over 2t^{2}} & {1-t^{2} \over 2t^{2}}\\
                           {1-t^{2} \over 2t^{2}} & {1+t^{2} \over 2t^{2}}
                        \end{array} \right),
\eeq
with the time-dependent dilaton $\Phi=\Phi_{0}+\ln t^{2}$, and the sizes of those cycles in $\tilde{x}$-space at $t=0$ are given by
\beq
           \tilde{s_{2}}^{2}\longrightarrow\infty,\qquad \tilde{s_{1}}^{2}\longrightarrow 8\pi^{2}.
\label{eq27}
\eeq
These are same as in $x$-space structure at $t=\infty$. Therefore, it would be convenient for $0<t\leq 1$ region to use $\tilde{x}$-space and for $(1\leq t<\infty)$ region to use $x$-space to describe physical phenomena. In fig.4 we show how the space of the solution I looks like as $t$ goes on.

\begin{center}
\leavevmode
\epsfysize=6cm \epsfbox{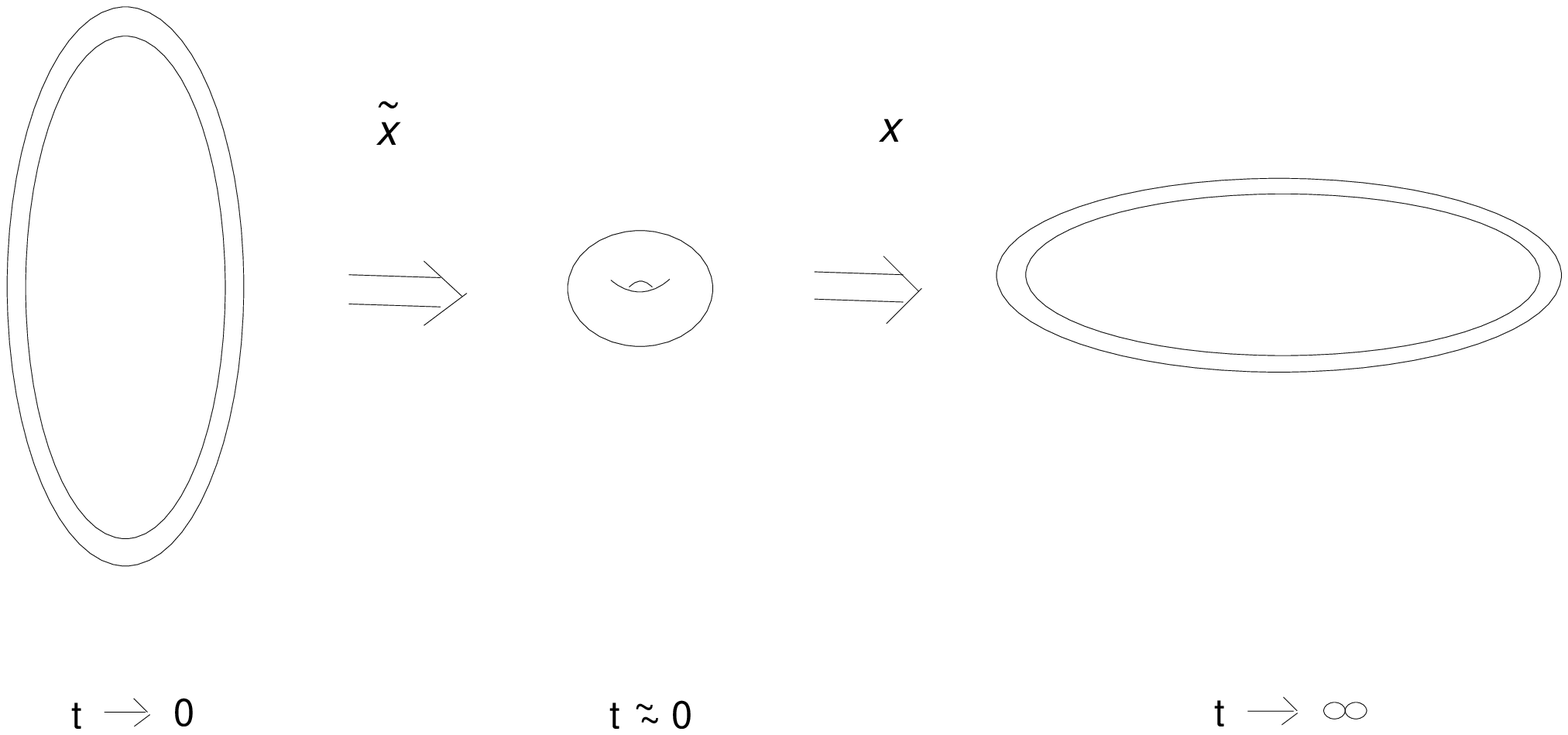}
\end{center}
\begin{center}
{\small fig.4~~}
\end{center}

Secondly, the difference of the lengths of cycles $s_{1}$ and $s_{2}$ in the asymptotic region $t=\infty$ has occurred even if the local metric is assumed to have common diagonal elements. This can be understood from fig.1-(a), where the parallelogram $OPQR$ moves toward $OP'Q'R$ and it collapses flat as $t$ goes to infinity, i.e. the distance $P'R$ approaches zero. However the scaling factor blows up as $t$ goes on $(\sqrt{-g}=t)$ and $s_{1}$ approaches a finite value $\sqrt{8}\pi$ as shown in (\ref{eq27}).

Thirdly, we have to consider the constraints required from the reparametrization invariance of the action (\ref{eq3}). The one is the vanishing Hamiltonian condition $\cite{giveon}$
\beqn
         H=-{1 \over 2}\: p_{0}^{~2}&+&{1 \over 2}\:[n_{i}(G^{-1})^{ij}n_{j}
\nonumber\\
               &~&\;+m^{i}(G-BG^{-1}B)_{ij}m^{j}+2m^{i}B_{ik}(G^{-1})^{kj}n_{j}]=0,
\label{eq28}
\eeqn
where only zero oscillation modes are chosen because we are interested in the low energy state modes. The integers $n_{i}$ and $m^{i}$ represent the momentum and the winding numbers, respectively. The equation (\ref{eq28}) determines the low energy spectra as
\beq
         p_{0}^{~2}=[n_{i}(G^{-1})^{ij}n_{j}
                         +m^{i}(G-BG^{-1}B)_{ij}m^{j}+2m^{i}B_{ik}(G^{-1})^{kj}n_{j}].
\eeq
Another constraint comes from the rescaling invariance $\int_{0}^{2\pi}d\sigma PX'=0$, which is equivalent to $\cite{giveon}$
\beq
         \sum_{i=1,2}n_{i}m^{i}=~0.
\label{eq30}
\eeq

For the solution I with $b=0$, for instance, the low energy spectrum is given by
\beqn
          p_{0}^{~2}&=&{1 \over 2t^{2}}~[(1+t^{2})n_{1}^{2}+2(1-t^{2})n_{1}n_{2}+(1+t^{2})n_{2}^{2}
\nonumber\\
                 &\qquad +&t^{2}\{(1+t^{2})(m^{1})^{2}-2(1-t^{2})m^{1}m^{2}+(1+t^{2})(m^{2})^{2}\}].
\label{eq31}
\eeqn
In (\ref{eq31}) as $t\rightarrow\infty$, the winding mode terms for nonvanishing $m^{i}$ diverge and so they are not excited to give $m^{1}=m^{2}=0$. The constraint (\ref{eq30}) is now automatically satisfied. The spectra in the asymptotic region is therefore given by
\beq
          p_{0}^{~2}\sim {1 \over 2}(n_{1}-n_{2})^{2}+{1 \over 2t^{2}}(n_{1}+n_{2})^{2},
\label{eq32}
\eeq
where $n_{1}'\equiv (n_{1}-n_{2})$ and $n_{2}'\equiv (n_{1}+n_{2})$ are the momentum excitation numbers in the asymptotic region, which are respectively associated with the cycles $s_{1}$ and $s_{2}$. The result, that the second term in (\ref{eq32}) provides dense spectra while the first term discrete ones in the asymptotic $t$-region, agrees with the results (\ref{eq24}) and (\ref{eq25}).

The similar argument works for $t\approx 0$ as well. Corresponding to the two momentum modes the effective dimensions of the asymptotic space is concluded to be $1+1$ both in $t\approx 0$ and $t\rightarrow \infty$.

In finite time regions, around $t\approx 1$ for example, the only constraint we have to consider is (\ref{eq30}) and energy spectra are determined by three independent quantum numbers. The result is natural because on such a small compactified torus both winding modes and center-of-mass momentum modes co-exist without  extreme high energy excitations. This fact has been recently analyzed in connection with the possibility of the information transmission from $t<1$ to $t>1$ region $\cite{hotta}$.


\section{The Solution II}

\begin{center}
\leavevmode
\epsfysize=6cm \epsfbox{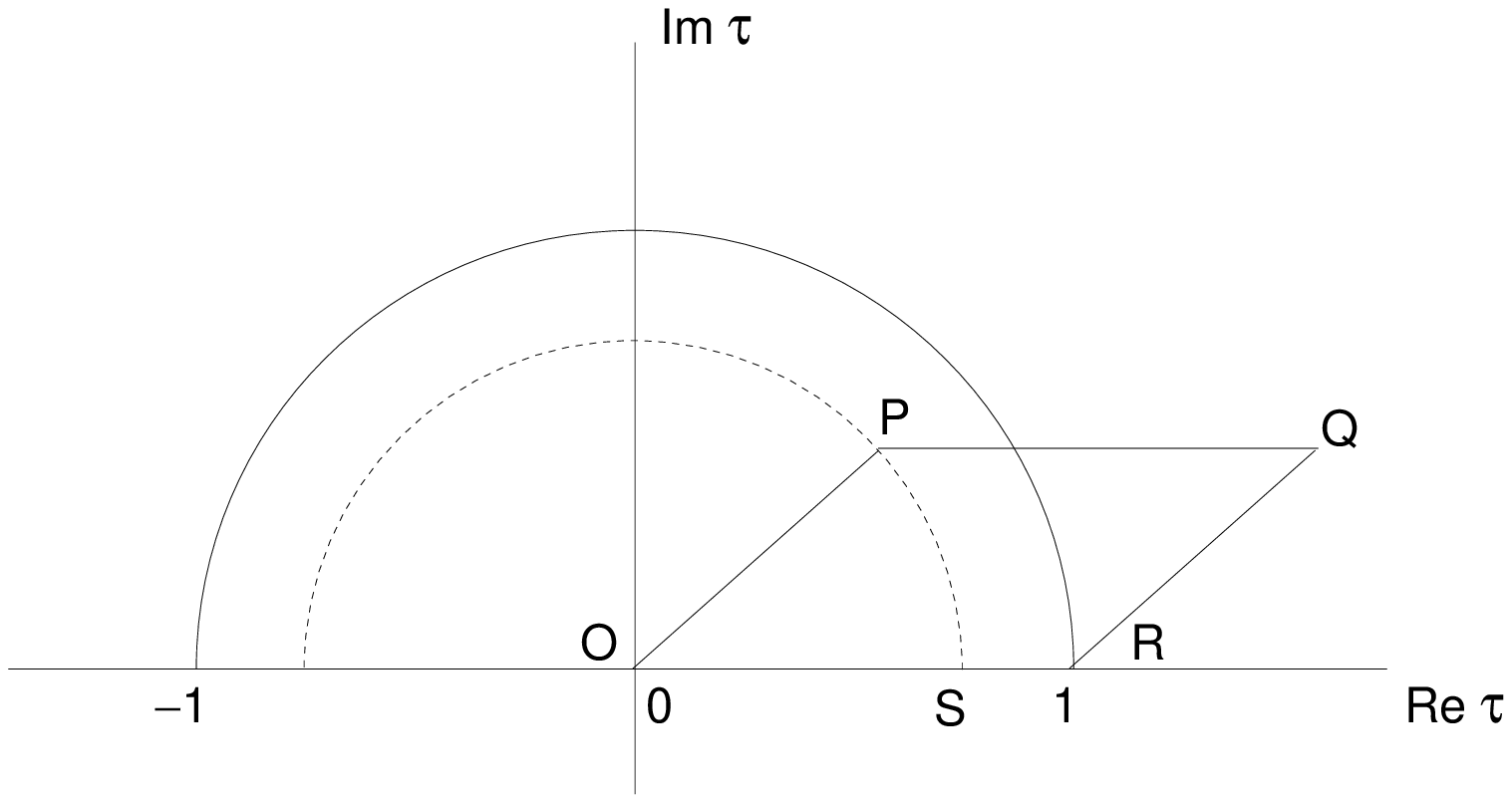}
\end{center}
\begin{center}
{\small fig.5 ~The motion of $\tau$ for the solution II.}
\end{center}

This is the case for the general solution under the condition of $g_{i0}=g_{0i}=0~(i=1,2)$ and $\Phi=const$. The metric solution we study here is given by (\ref{eq15}) with $b=0$ in (\ref{eq16}). The space structure can be represented with the parallelogram OPQR on $\tau$-plane in fig.5, and as $t$ grows the point P moves toward S along a quasi-semicircle, shown with a dotted line in fig.5. The path is deformed from a semi circle because of the time dependence of $e^{\beta}$. In this case $\overline{OP}\ne\overline{OR}$. The solution I is a special case where $\overline{OP}=\overline{OR}$. One of the cycles of the space is chosen along OR and the length is given by (\ref{eq23}):
\beq
        s_{2}^{~2}= 4\pi^{2}e^{-\beta}\sqrt{t^{2}+c^{2}\left({t^{2} \over 2}+d\right)^{2}}.
\eeq
The other cycle $s_{1}$, as in the case of (\ref{eq25}), is chosen among the other topologically independent cycles in such a way that the length is shortest. For small value of $t$, $\overline{OQ}$ and then $\overline{OP}$ are legitimate and the lengths are respectively given by
\beq
        s_{1}^{~2}=\left\{ \begin{array}{l}
                            \overline{OQ}^{2}=4\pi^{2}\left\{(e^{\beta}+e^{-\beta})\sqrt{t^{2}+c^{2}\left({t^{2} \over 2}+d\right)^{2}}+2c\left({t^{2} \over 2}+d\right)\right\}\\
                            \overline{OP}^{2}=4\pi^{2}e^{-\beta}\sqrt{t^{2}+c^{2}\left({t^{2} \over 2}+d\right)^{2}}\\
                            \qquad\quad\cdots\cdots\cdots
                             \end{array} \right.
\eeq

In order to understand the space structure in the asymptotic region where $t\rightarrow\infty$, we first consider the case of $\overline{OP}\:/\:\overline{OR}$ being a rational number, namely,
\beq
        {\overline{OP} \over \overline{OR}}=\sqrt{{G_{11} \over G_{22}}}=e^{\beta}\rightarrow~{l \over k}~\q (t\rightarrow\infty),
\eeq
provided $k$ and $l$ being positive and mutually prime integers. In this case one can show that the length of the shortest cycle $s_{1}$ turns out to be
\beqn
        s_{1}^{~2}&=&8\pi^{2}ke^{\beta}\left\{\sqrt{t^{2}+c^{2}\left({t^{2} \over 2}+d\right)^{2}}-c\left({t^{2} \over 2}+d\right)\right\}
\nonumber\\
                  &\longrightarrow&{8\pi^{2}l \over c}~\q (t\rightarrow\infty),
\label{eq36}
\eeqn
which is finite. To understand this result, let us look at the case where $k=3$ and $l=2$ (fig.6), where the space is represented on a lattice, each block of which represents our space. As the time $t$ grows, the angle $\angle~PO'R'$ collapses to zero and the point $P$ approaches to $R'$. For large $t$, therefore, the distance $\overline{PR'}$ is shorter than any other cycles although it crosses $s_{2}$-cycle three times. Taking account of the scale factor $\sqrt{-g}=t$, one can confirm $s_{1}$ approaching to a finite value (\ref{eq36}) with $l=2$. For other rational values of $e^{\beta}=l/k$, (\ref{eq36}) may be easily confirmed.

\begin{center}
\leavevmode
\epsfysize=5cm \epsfbox{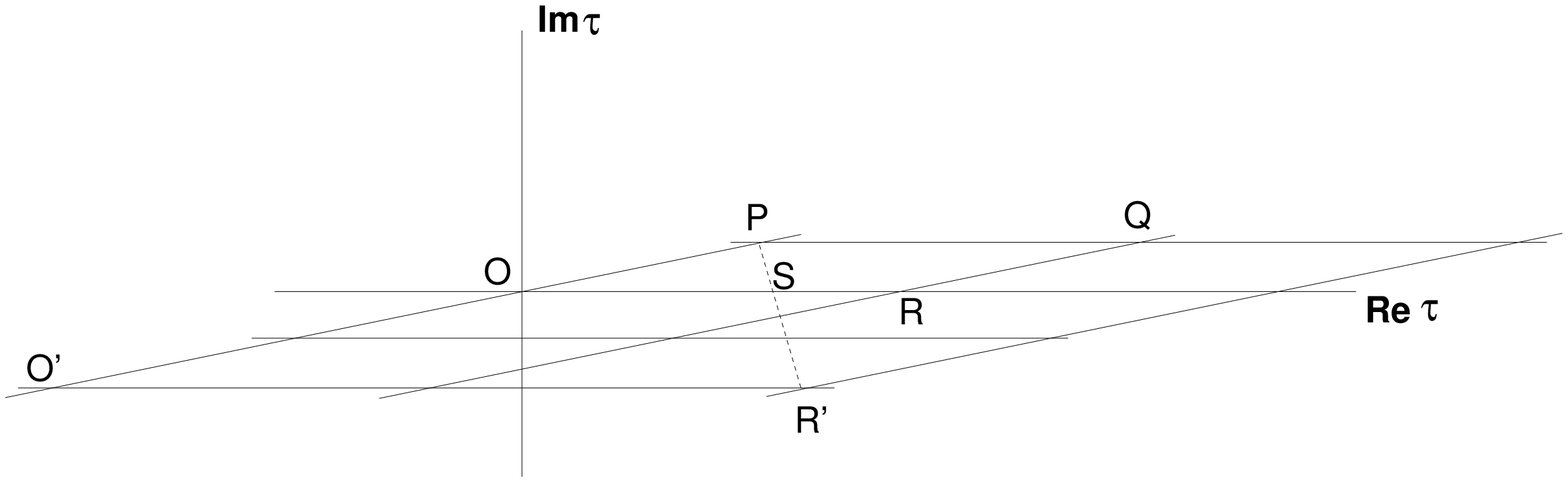}
\end{center}
\begin{center}
{\small fig.6 ~The shortest cycle in the limit of $t\rightarrow\infty$.}
\end{center}

On the other hand, another cycle $s_{2}$ grows in proportion to $t$ because the space is again stretched in one direction while the other cycle kept finite.

It may be inferred that, although the cycle $s_{1}$ crosses the other cycle $s_{2}$ three times, there may be another choice of $s_{2}'$ such that the cycle $s_{1}$ crosses the new cycle $s_{2}'$ only  once. The cycle $\overline{PR}$ in fig.6 is the one but the length grows to infinity again.

In case of the value $\sqrt{G_{11}/G_{22}}=e^{\beta}$ being an irrational number, one can choose a series of rational numbers, $l/k$, which converges to the irrational number $e^{\beta}$ as $l\rightarrow\infty$. The length $s_{1}$, as is observed in (\ref{eq36}), grows to infinity as well as $s_{2}$. It should however be, noticed that the path $s_{1}$ crosses $s_{2}$ infinite times.

To look at the space around $t\approx 0$, we again study the $T_{D}$-transformed solution. The metric is now given by
\beq
         G_{D}={1 \over t^{2}}\left( \begin{array}{cc}
                 e^{-\beta}\sqrt{t^{2}+c^{2}\left({t^{2} \over 2}+d\right)^{2}} & -c\left({t^{2} \over 2}+d\right) \\
                 -c\left({t^{2} \over 2}+d\right) & e^{\beta}\sqrt{t^{2}+c^{2}\left({t^{2} \over 2}+d\right)^{2}} \\
                        \end{array} \right),
\eeq
with the dilaton
\beq
        \Phi=\Phi_{0}+\ln t^{2},
\eeq
and the cycle sizes are generally given by
\beqn
         &\tilde{s_{1}}^{~2}&:~~\infty~\longrightarrow~2\pi^{2}c\left\{\left(e^{\beta}+e^{-\beta}\right)-2\right\}
\nonumber\\
\\
         &\tilde{s_{2}}^{~2}&:~~\infty~\longrightarrow\qquad~ 2\pi^{2}e^{\beta}c.
\nonumber
\eeqn
The reason why both of $\tilde{s_{1}}$ and $\tilde{s_{2}}$ are finite as $t\rightarrow \infty$ comes from the fact that the upper side of the parallelogram is shifted along the cycle $\tilde{s_{2}}$ so that the closed cycle $\tilde{s_{1}}$ must also be stretched along the $\tilde{s_{2}}$ direction. If one diagonalizes the metric $G_{D}$ in the limit $t\rightarrow \infty$ one of the eigenvalues approaches to zero while the other to a finite value, hence the torus actually collapses to a one-dimensional ring with a finite length in the limit. This is consistent with the fact that the area of torus approaches to zero as the scale factor $\sqrt{-g_{D}}=1/t$ shows.

Finally, let us look at the energy spectra coming from the zero modes. The energy $p_{0}$ is given by
\beqn
         p_{0}^{~2}&=&{1 \over t^{2}}\left[e^{-\beta}\sqrt{t^{2}+c^{2}\left({t^{2} \over 2}+d\right)^{2}}n_{1}^{~2}-2c\left({t^{2} \over 2}+d\right)n_{1}n_{2} \right.
\nonumber\\
          &+&e^{\beta}\sqrt{t^{2}+c^{2}\left({t^{2} \over 2}+d\right)^{2}}n_{2}^{~2}+t^{2}\left\{e^{\beta}\sqrt{t^{2}+c^{2}\left({t^{2} \over 2}+d\right)^{2}}(m^{1})^{~2} \right.
\nonumber\\
          &+& \left.\left. 2c\left({t^{2} \over 2}+d\right)m^{1}m^{2}+e^{-\beta}\sqrt{t^{2}+c^{2}\left({t^{2} \over 2}+d\right)^{2}}(m^{2})^{~2}\right\}\right],
\eeqn
which approaches to
\beq
         {c \over 2}\left(e^{{\beta \over 2}}m^{1}+e^{-{\beta \over 2}}m^{2}\right)^{2}t^{2}+(finite~terms~as~t\rightarrow\infty),
\eeq
as $t$ goes to infinity. Since infinite excitation energy should be suppressed, one has to again choose $m_{1}=m_{2}=0$ for large $t$, which satisfies the constraint (\ref{eq30}) as well. The energy spectra in this asymptotic region are now given by
\beqn
        p_{0}^{~2}&=&{c \over 2}\left(1+{2cd \over t^{2}}\right)\left(e^{-{\beta \over 2}}n_{1}-e^{{\beta \over 2}}n_{2}\right)^{2}
\nonumber\\
                  &~&+{1 \over ct^{2}}\left(e^{-\beta}n_{1}^{~2}+e^{\beta}n_{2}^{~2}\right).
\label{eq42}
\eeqn

If $e^{\beta}$ is a rational number, the first term in (\ref{eq42}) provides a series of discrete  spectra, while the second term a series of continuous ones. The result is consistent with the fact that $s_{1}$ is finite while $s_{2}$ grows infinity as $t\rightarrow\infty$.

In the case of $e^{\beta}$ being an irrational number, there exists a series of numbers $n_{1}/n_{2}$ which accumulates to $e^{\beta}$, and even the first term provides a series of continuous spectra. This is again consistent with the cycle lengths discussed before.

As for the region where $t\approx 0$, the situation is same as above except that now the winding modes survive instead of the momentum modes.


\section{Conclusion}

We have studied the time development of two-dimensional string universe with torus structures. If the ratio of the two parameters $\sqrt{G_{11}/G_{22}}=e^{\beta}$ approaches to a rational number as $t\rightarrow\infty$, the space has been shown to be asymptotically stretched out along a cycle, while the length of the other cycle is kept finite at around a Planck size. This suggests to us an interesting possibility that even in the critical string theories, some characters of the metric might imply the shrinkage of a certain number of dimensions of the space.

For irrational values of $e^{\beta}$, the sizes of two cycles and also the space volume (2-d area) grow large in proportion to the proper time $t$ in the asymptotic region.

If one traces the expanding universe backward in time to the past, the space shrinks to the space having both dimensions finite, and enters to the universe where $\tilde{x}$ space dominates. The space again grows large as $t$ approaches to zero. No spatial singularity appears around $t\approx 1$ where the space size is minimum. In this model the large size universe in $\tilde{x}$-space comes into the world at $t\approx 0$ and then develops.

In connection with the shrinkage of partial dimensions Mueller $\cite{mueller}$ studied in the case of the space having $(S^{1})^{D-1}$ structure. In his model, if the radius of $i$-th circle, $r_{i}$, is given by
\beq
          r_{i}=\alpha_{i}(t-t_{0})^{p_{i}}~,\qquad ~i=1,2,\cdots,D-1.
\eeq
A condition in the asymptotic region
\beq
          \sum_{i=1}^{D-1}p_{i}^{2}=1,
\eeq
is obtained for real constants $p_{i}$. Our result is consistent with his condition because for $D=3$ we can choose $p_{1}=0$ and $p_{2}=1$.

In our work we have disregarded all the oscillation modes, so that the arguments cannot be applied to the cosmology where the high temperature effects may play a crucial role around $t\approx 1$. However, if we regard the two-dimensional space to be part of the critical string theory in the asymptotic time region, the present results are of great interest since the small sized compact space follows as a result of a certain condition over the metric tensor.

The authors want to thank M.Anazawa for discussions. The work is supported in part by the Grant-in-Aid for Scientific Research from the Ministry of Education, Science and Culture $(\sharp 08640372)$ and also in part by The Japan-Former Soviet Union Scientists Collaboration Program in Japan Society for the Promotion of Science (JSPS). The authors are grateful for these financial supports.




\end{document}